\documentclass[preprint,12pt]{elsarticle}

\usepackage{graphicx}
\usepackage{xcolor}
\usepackage{float}

\usepackage{amssymb}

\usepackage{tikz}
\usepackage{graphicx}

\usepackage{wasysym}
\usepackage{eucal}
\usepackage{amsfonts}
\usepackage[table,xcdraw]{}

\journal{Chaos, Solitons \& Fractals}

\begin{document}

\begin{frontmatter}


\title{Analytic approaches of the anomalous diffusion: \\ a review }



\author{Maike A. F. dos Santos*}

\address{*Physics Institute, \\ Federal University of Rio Grande do Sul,\\
Postcode 15051, CEP 91501-970,\\
Porto Alegre, RS, Brazil}

\begin{abstract}
This review article aims to stress and reunite some of the analytic formalism of the anomalous diffusive processes that have succeeded in their description. Also, it has the objective to discuss which of the new directions they have taken nowadays. The discussion is started by a brief historical report that starts with the studies of thermal machines and combines in  theories such as the statistical mechanics of Boltzmann-Gibbs and the Brownian Movement. In this scenario, in the twentieth century, a series of experiments were reported that were not described by the usual model of diffusion. Such experiments paved the way for deeper investigation into anomalous diffusion, i.e. $\langle (\Delta x)^2 \rangle \sim t^{\alpha}$ to $\alpha \neq 1$. These processes are very abundant in physics, and the mechanisms for them to occur are diverse. For this reason, there are many possible ways of modelling the diffusive processes. This article discusses three analytic approaches to investigate anomalous diffusion: fractional diffusion equation, nonlinear diffusion equation and Langevin equation in the presence of fractional, coloured or multiplicative noises. All these formalisms  presented different degrees of complexity and for this reason, they have succeeded in describing anomalous diffusion phenomena.
\end{abstract}


\end{frontmatter}



\newpage

\section{Introduction}

Nowadays, we are all sure of the existence of atoms and molecules, but it has not always been so. The search for the understanding of this microscopic world has involved hundreds of theories that have direct implications in our lives. One theory that investigated the effects macroscopic effects of a system of many particles (gas) and was very successful, was the thermodynamics.
The phenomenological theory known as thermodynamics sought to describe the phenomena related to heat. This theory obtained great strides in the investigations of Nicolas L\'eonard Sadi Carnot (1796-1832) and Rudolf Julius Emanuel Clausius (1822-1888) during the XIX century. At that time, the hypothesis of the atomic and molecular world was in evidence, in which physicists Ludwig Boltzmann (1844-1906) and Josiah Gibbs (1839-1903) led physicists to investigate the thermodynamic quantities from a molecular and atomic point of view. In other words, these scientists of nature have discovered several laws that nature apparently obeys. 

The molecular and atomic world began to gain his own mathematical approaches. 
In this context, James Clerk Maxwell (1831-1879) proposes a distribution associated with a thermodynamic system in equilibrium. Such a distribution has a Gaussian shape at velocities, which would imply satisfying some important properties associated with the velocity of the particles of a gas. In 1855, Adolf Eugen Fick (1829-1901) derived his laws, which relate the flow of particles in a volume with the temporal variation of concentration in volume. In particular, he obtained an equation describing the dynamics of a particle concentration in space and time, which mathematically has the structure of the heat equation. These formalisms described the dynamic processes in a phenomenological way, and not the mechanisms that imply in the diffusion.

The ideas about the microscopic dynamics implied in a series of works done by Boltzmann and Gibbs, that investigated the microscopic aspects of the thermodynamics, thus discovering a connection between the macroscopic world and the microscopic one, founding the statistical mechanics. In this context, the Boltzmann-Gibbs entropic function is presented, i.e. 
$ S_{BG}=-k_B \sum_{i=1}^{W} p_i \log p_i$, 
in which $k_B$ is the Boltzmann's constant and $p_i$ is the probability associated with a microstate of the system. In fact, Boltzmann brought a probabilistic character to all the mechanics of particles system  in thermodynamics. The result of this discovery were diverse. In particular, it implies several analytic expressions to describe thermodynamic systems. In these circumstances, the Boltzmann-Gibbs entropy can be related to the Maxwell distribution for the velocity of the particles, that is to say, a function of the Gaussian type. The diversity of problems that could be described by statistical mechanics \cite{boltzmann1970weitere,Gibbs,tolman} transcended physics and, today, has developments in the study of complex systems \cite{auyang1999foundations}. But there are a number of complex processes in which the mechanics of Boltzmann and Gibbs are not applicable, for example, in processes dynamics with memory between their components, in such a way that ergodic hypothesis is violated \cite{walters2000introduction,deng2009ergodic,rebenshtok2008weakly}.  In this scenario, the investigations about microscopic world has begun to have a probabilistic approach, a series of formalism associated to the dynamics of small particles were proposed.  Lately,  the formalisms defined the concept of usual diffusion, the first formalism was proposed by Einstein in the beginning of twentieth century.

In 1905, Albert Einstein (1879-1955) published a series of works that had a great impact on modern physics, among which is the one that described the movement of microscopic particles (now known as Brownian motion) immersed in a fluid \cite{einstein1}. In his theory, he proposed a diffusive system without temporal and spatial correlations between its components.  Thereby, he derived the diffusion equation. Through this equation he showed which the fluctuation of position  to particles is associated with a Gaussian distribution and the temporal evolution implies a law of proportion for the mean square displacement (MSD), given by
\begin{eqnarray}
\left \langle(x- \langle x \rangle )^2
\right \rangle  \propto t. \nonumber
\end{eqnarray}
This behaviour is connected to Fokker-Planck equation (diffusion equation under presence of force). Moreover, the Fokker-Planck equation is deeply connected with  Boltzmann-Gibbs entropy \cite{PhysRevE.91.042140,chavanis2008nonlinear,htheorem,risken}. After Einstein's proposal, several physical systems have demonstrated to follow this ratio for the time evolution of quadratic displacement, on the other hand: a number of physical systems broke the relationship of MSD in time, i.e., $\langle(x- \langle x \rangle )^2 \rangle \neq t$, cases that became known as anomalous diffusion phenomena (Non-Brownian motion) \cite{ctrw}.

Non-Brownian behaviour in most cases is associated with generalised distributions and, therefore, more general theories than Boltzmann-Gibbs mechanics. Thus, several physical mechanisms could cause this type of diffusion, which was characterised as anomalous diffusion. Among the most well known mathematical formalisms are nonlinear dynamics, fractional equations (fractional calculus \cite{podlubny1998fractional}), and equations with variable diffusion coefficients. These formalisms can be treated analytically and derive a MSD of the type $\langle(x- \langle x \rangle )^2 \rangle \propto t^{\nu}$, which may imply minimalist generalisations of the formalisms that describe the usual diffusive movement, for $\nu \rightarrow 1$ the usual case is recovered, to $\nu>1$ systems are known to have a super-diffusive dynamic and to $\nu<1$ sub-diffusive.

The fractional diffusion involves phenomena that have spatial and temporal correlations. In this work, it is common the nomenclature \textit{fractional dynamics} that refers to diffusion equations defined by derivatives of non-integer order (the derivative can assume a value of real order). Anomalous diffusion through fractional equations is associated with  super--statistics \cite{chechkin2017brownian} and can be linked to a generalised random walk, which has implications for animal dynamics \cite{alves2016transient}, in the movement of proteins \cite{schutz1997single} and even in the diffusion of magnetic particles from the sun \cite{qsol}.

The nonlinear diffusion equations imply in a series of generalisations, in particular, the nonlinear dynamics can imply in anomalous diffusive processes and in generalised entropies \cite{htheorem1,dos2017nonlinear}. These entropic forms recover the Boltzmann-Gibbs entropy, so the type of treatment to describe the anomalous phenomena is associated with generalised statistics. In particular, these generalised entropic forms have applications in various contexts and are associated both in the thermodynamic description of massive objects such as black holes \cite{tsallis2013black} as to quantum characteristics involving the dynamics of spins in glasses \cite{qspin}. 

The formalism that addresses the problem of anomalous diffusion for diffusion equations with variable diffusion coefficients are associated with the heterogeneous media in which the diffusion coefficient can be time and space dependent. Such formalism is associated with the Boltzmann-Gibbs entropy, but it is restricted to a class of solutions that is related to stochastic processes with a multiplicative noise type in Langevin sense \cite{langevin1}. This type of process can justify the dynamics associated with out-of-equilibrium thermodynamics \cite{entropyproduction1}. The variable diffusion coefficient approach can be connected to a class of nonlinear equations associated with generalisations of statistical mechanics \cite{lisa1}.

Several generalisations with respect to the Boltzmann-Gibbs theory have been proposed, from the dynamic point of view, and these generalisations describe anomalous diffusion phenomena. In this text, we will discuss the formal aspects that led to the description of anomalous diffusion. This work is divided into a few sections. In the  section \ref{sec1}, we will review the first analytic proposal to describe an anomalous diffusive process in turbulent currents. In the same section, we will make a brief classification on anomalous diffusion. In the others sections, we will review three analytic approaches to anomalous diffusion. The first of them, section \ref{sec2} concerns the construction of walkers that imply in equations of diffusion with fractional derivatives i.e. $\frac{d^{\alpha}\ }{d x^{\alpha}}$ \cite{podlubny1998fractional}. A typical example is the following expression:
\begin{eqnarray}
\frac{\partial^{\alpha}\rho}{\partial t^{\alpha}}  = \mathcal{D}_{\alpha,\mu} \frac{\partial^{\mu} \rho}{\partial x^{\mu}}. 
\end{eqnarray}
in which $\rho(x,t)$ is a distribution of probabilities, $\mathcal{D}_{\gamma,\mu}$ is the generalised diffusion coefficient, $0<\alpha<1$ and $1<\mu<2$. This equation will be detailed throughout the text, this formalism gained much evidence with the works of Klafter, Barkai, Saichev
Zaslavsky, 
Schneider,
 Wyss and Metzler \cite{ctrw,PhysRevLett.82.3563,schneider1989fractional,saichev1997fractional}.  In the second analytic method (section \ref{sec3}), we demonstrate in what context non-linear diffusion equations \cite{vazquez2007porous,nonlinear} appear in physics, i.e.
\begin{eqnarray}
\frac{\partial \rho}{\partial t}  = \mathcal{D}_{\gamma} \frac{\partial^2 \rho^{\gamma}}{\partial x^2}. 
\end{eqnarray}
 and how do these forms connect with the theory proposed by Tsallis about a possible generalisation of statistical mechanics \cite{tsallis}. The last formalism (section \ref{sec4}) is about stochastic equations. Such equations admits a series of generalisations,   among them we may cite the generalised Langevin equation \cite{kubo,mandelbrot,langevin3}, as follows
 \begin{eqnarray}
m\frac{d^2}{dt^2}x(t)=- \varphi \int_{0}^t dt'  \zeta(t-t') \frac{d}{dt'}x(t') \ + \eta \cdot \xi(t), 
\label{eq001}
\end{eqnarray}
in that $x(t)$ is a stochastic variable dependent of time. The Eq. (\ref{eq001}) describes the position of a particle of mass $m$ immersed in a fluid. There, $\zeta(t)$ is the correlation function and $\xi(t)$ the noise. We will review a series of details of these three formalisms which reproduce very well a series of dynamic behaviours in diffusive systems.
 
Finally, in the section \ref{sec5} we will discuss how these formalisms have been used, and how their applications transcended their initial purpose.  Looking at the big picture of the diffusive process studies we observe that nature is not simple and involves a diversity of complex factors. In this scenario, the analytic tools become elegant because these tools capture the sophisticated dynamic behaviour in nature.

\section{Turbulent diffusion and the beginning of anomalous diffusion}
\label{sec1}

The anomalous diffusion processes began to be observed and investigated after  Einstein's work \cite{einstein1}. The year 1928 was a milestone in the history of anomalous diffusion, in this year was published the work \textit{''Atmospheric Diffusion shown on a Distance-Neighbour Graph''} \cite{richardson1}, written by the British scientist Lewis Fry Richardson, it is based on measures of diffusive  $D$ in the turbulent systems, ranging from capillary tubes ($D \sim 10^{-2}\frac{cm^2}{sec}$) up to cyclones ($D \sim 10^{11} \frac{cm^2}{sec}$). To understand this variation of diffusivity, Richardson demonstrated that the equation proposed by Fick for diffusion, i.e.,
\begin{eqnarray}
\frac{\partial \rho}{\partial t}=D\frac{\partial^2 \rho}{\partial x^2},
\end{eqnarray}
is not suitable for describing diffusion in turbulent currents, in which $\rho$ is concentration of particle. Thus, Richardson proposed the number of neighbours per unit length $q$ as a function of the distance $l$ between neighbours \cite{richardson1}. And he obtained the equation he called \textit{"Non-Fickian Diffusion equation"}, 
\begin{eqnarray}
\frac{\partial \overline{q}}{\partial t}=\frac{\partial }{\partial l}\Big(F(l)\frac{\partial \overline{q}}{\partial l} \Big),
\end{eqnarray}
in which $F(l)$ is  increasing in relation to $l$. He treated the problem as having the diffusive independent of position (spacial variable), but dependent on the distance between particles. To determine the function $F(l)$, Richardson constructed graphs of the respective logarithms of the diffusivity $D$ versus the distance between the neighbours $l$.

In this way, he adjusted the data through the relation $F(l)=0.2 l^{\frac{4}{3}}$, then obtaining the equation
\begin{eqnarray}
\frac{\partial \overline{q}}{\partial t}=\epsilon\frac{\partial }{\partial l}\Big(l^{\frac{4}{3}}\frac{\partial \overline{q}}{\partial l} \Big).
\label{rich1}
\end{eqnarray}
Considering the change of variable
 $\alpha=l^{\frac{1}{3}}$, we can rewrite the Eq. (\ref{rich1}) as follow
\begin{eqnarray}
\frac{\partial \overline{q}}{\partial t}=\frac{\epsilon}{9}\Big(\frac{2}{\alpha}\frac{\partial \overline{q}}{\partial \alpha}+\frac{\partial^2 \overline{q}}{\partial \alpha^2}\Big),
\label{richardsonl}
\end{eqnarray}
which has the same mathematical structure of the heat diffusion equation in a homogeneous solid, where the isothermal surfaces are concentric spheres of radius $\alpha$ and with equal diffusivity. Richardson found the solution to Eq. (\ref{richardsonl}), as
\begin{eqnarray}
\overline{q}(\alpha,t)=A(4t \epsilon)^{-\frac{3}{2}}e^{-\frac{\alpha^2}{4 t \epsilon/9}},
\end{eqnarray}
in which $A$ is an independent parameter of $\alpha$ and $t$.  Thus, Richardson  defined the normalisation constant as follows
\begin{eqnarray}
\mathcal{N} &=&\int_{-\infty}^{+\infty}\overline{q} dl \nonumber \\ &=& 3A \int_{-\infty}^{+\infty} e^{-\beta^2}\beta^2 d\beta
\end{eqnarray}
in which $\beta^2=\frac{\alpha^2}{4 t \epsilon/9}$. He obtained the second moment, given by
\begin{eqnarray}
\langle l^2 \rangle =\frac{105}{16}(4t \epsilon/9)^3\mathcal{N}.
\label{richarsonmsd}
\end{eqnarray}
This expression demonstrates a different behaviour in relation to Fickian diffusion, which would later be classified as anomalous, and the diffusive processes that involved this class of phenomena would be known as \textit{anomalous diffusion}. 

In the decades following Richardson's work, some experimental work emerged confirming the existence of this type of behaviour. Let's mention some examples: In polymers \cite{polimeros1,polimeros2}, plasma \cite{plasma1,plasma2}, metals
 \cite{metais1,metais2} and semiconductors \cite{semicondutores1}. Despite strong experimental evidence on the subject, the formalisms for describing such phenomena were unknown.

In the second half of the last century, some formalisms have arisen to describe this type of transport with average nonlinear quadratic displacement over time. The anomalous diffusive processes came to be classified in two ways, to define them we will consider the equation 
\begin{eqnarray}
\langle (x - \langle x\rangle )^2 \rangle \propto t^{\alpha}.
\end{eqnarray}
The first class is that of super-diffusive phenomena, which occur when $\alpha >1$. The second class is that of sub-diffusion (sub--diffusive phenomena), which occurs to $\alpha<1$, the case when  $\alpha =1$ corresponds to the usual diffusion. In 1975, Scher and Montroll \cite{scher1975anomalous,montroll1973random} related the anomalous diffusion in the dispersive transport of charge carrier  in amorphous semiconductors, as the investigation of continuous time random walk (CTRW) \cite{montroll1965random}  approached by Weiss and Montroll. To a deeper comprehension about anomalous diffusion in fractional context see the references \cite{bouchaud1990anomalous,metzler2014anomalous,zaburdaev2015levy,mainardi2001fundamental}.

 These systems that present anomalous behaviour can be found on diffusion of proteins in the cellular membrane \cite{proteinas1,proteinas2}, hydrology and geology \cite{geology1,hydrodynamic1},  among others \cite{ctrw}. Super-diffusive behaviours were found in the movement (corresponding to processes known as L\'evy's walk) of various animal species \cite{giuggioli2010animal,macacoaranha}. It has also been used at a microscopic level to describe the movement of protozoa \cite{alves2016transient}. In chemistry, anomalous diffusion has been a phenomenon that occurs in systems with chemical reactions or in electrical impedance theory \cite{impedancia}, among other applications that will be mentioned throughout the text. In the following sections, we will discuss the formalisms that imply in anomalous diffusion.

\section{Fractional equations and continuous time random walk}
\label{sec2}

In the case of the random continuous walker known as CTRW \cite{ctrw}, unlike the previous case, the time interval between the steps is given by an infinitesimal amount of time. The walker takes a long step $\lambda$ in an arbitrary direction in $x$, for a given time interval between $t$ and $t+dt$. The steps are statistically independent, occurring at random time intervals.
We can then write the length of the jump as a probability density function,
\begin{eqnarray}
\lambda (x) = \int_0^\infty \psi (x, t) dt ,
\end{eqnarray}
\noindent as well as the waiting time
\begin{eqnarray}
w(t) = \int_{-\infty}^\infty \psi (x, t) dx ,
\end{eqnarray}
\noindent in which $\lambda (x) dx$ corresponds to the probability of a long jump $L$ in a given range $x \to x + dx$, and $w(t) dt$ the probability of a waiting time $\tau$ in a time interval $t \to t+dt$. Thus, the walker can be described by a probability density function $\psi (x, t)$, being $L$ and $\tau$ independent random variables. The function  $\psi (x, t)$ can be decoupled as follows $\psi (x, t) = w(t) \lambda(x)$.
In this way, there may be a divergence in both the waiting time and the length of the jumps, depending on the nature of the functions $\omega (t)$ and $\lambda (t)$. These quantities may describe characteristics of the distribution. For example, the case with finite mean wait time and divergent jump length variance implies L\'evy-type distributions, or else the case where the average wait time diverges, keeping the jump length variance constant, implies in the random walker with fractal time.

Thus, a parallel between random walkers with discrete and continuous time is established in the case where time is a discrete variable. There are successive jumps occurring between uniform time intervals, but, in case that time continuously evolves, the duration between jumps constitutes the random variable. In this way, the prediction of the walker's next position may not only want local knowledge of walking but also of positions in earlier times. This dependence on the state of the system and its past history reveals that the CTRW  can describe a non-Markovian process.

The diffusion equation can be obtained by means of the integral equation of CTRW theory, obtained by means of the Fourier transform, following procedure. It is considered the average waiting time,
\begin{eqnarray}
\tau = \int_0^\infty dt w(t) t  ,
\end{eqnarray}
and the jump length variance
\begin{eqnarray}
\sigma^2 = \int_{-\infty}^{\infty}dx \lambda(x) x^2 .
\end{eqnarray}
By means of such averages, we can characterise different types of CTRW  considering the finite or divergent nature of these quantities. In a more general case, any of these different CTRW can be described by the integral equation
\begin{eqnarray}
\eta(x, t) = \int_{-\infty}^\infty dx' \; \int_0^t dt' \eta(x', t') \psi(x-x', t-t') + \delta(x) \delta(t),
\label{11,2,3}
\end{eqnarray}
 being  $\eta(x, t)$ the probability per unit of displacement and time of a random hiker who has left the $x$ in the time $t$, to the position $x'$ in time $t'$, being the last term (product of two delta functions) the initial condition of the walker.

Therefore, the probability density function $p(x, t)$ of the walker to be found in $x$ in the time $t$ is given by
\begin{eqnarray}
p(x, t) = \int_0^t dt' \eta(x, t') \; \Phi(t-t') ,
\label{11,2,4}
\end{eqnarray}
in which
\begin{eqnarray}
\Phi(t) = 1- \int_0^t dt' w(t'),
\label{11,2,5}
\end{eqnarray}
 is the probability of the walker not jumping during the time interval $(0,t)$, that is, to remain in the initial position. Applying the Laplace transform in equations (\ref{11,2,4}) and  (\ref{11,2,5}) and using the convolution theorem, we have
\begin{eqnarray}
p(x, s) = \frac{1}{s} \eta(x, s) [1-w(s)] .
\label{11,2,7}
\end{eqnarray}
To determinate $\eta(x,s)$, we must return to (\ref{11,2,3}) and apply the Laplace transform on the temporal variable and Fourier transform  on the spatial variable.  Making use of integral transformations, we have
\begin{eqnarray}
\eta(k, s) [1-\psi(k, s)]=1 .
\end{eqnarray}
Using the previous result (Eq. (\ref{11,2,7})) and considering a generic initial condition $p_0 (x)$, we have
\begin{eqnarray}
p (k, s) = \frac{1-w(s)}{s}\frac{p_0 (k)}{1-\psi(k, s)} .
\label{distributionctrw}
\end{eqnarray}
\noindent This equation can be applied to systems that have the jump length coupled to the waiting time.

\subsection{The walker associated with fractional equation}

 In 1987, Klafter, Blumen and Shlesinger \cite{klafter} demonstrated how it is possible to arrive at anomalous diffusive behaviours starting from the continuous random walk in time. Initially, if we consider that $\langle x^2 \rangle$  is finite however $\langle t \rangle$ infinite (divergent), the waiting time distribution  assumes long-tailed with power-law asymptotic behaviour
\begin{eqnarray}
 w(t) \sim  \left(\frac{\tau}{t}\right)^{1-\alpha},
 \label{ctrww}
\end{eqnarray}
 which has the corresponding Laplace space $ w(s) \sim 1 -  (\tau s)^{\alpha}$. As  $\lambda$-distribution is finite
 \begin{eqnarray}
\lambda(x)=(2 \pi \sigma^2)^{-\frac{1}{2}}\exp \left[-\frac{x^2}{2 \sigma^2}\right],
\label{ctrwalb}
\end{eqnarray}
 with the asymptotic limit $\lambda(k)\sim 1-\sigma^2 k^2/2$.

\begin{figure}[h]
\centering
\includegraphics[scale=1.0]{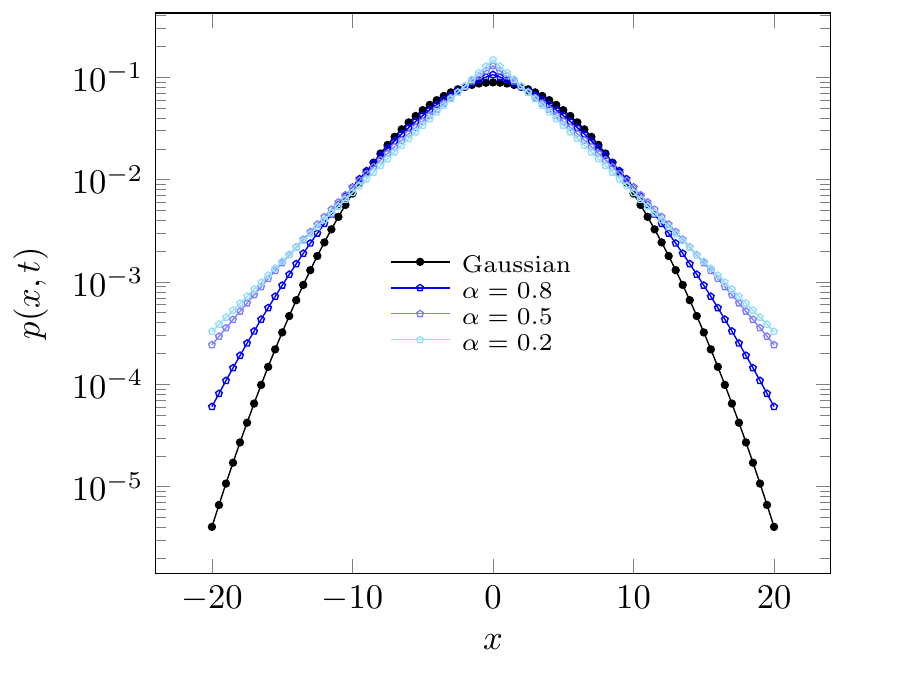}
\caption{\small{The figure present the  probability distribution for different values of $\alpha$ parameter.  We consider $\mathcal{K}_{\alpha}=10$ and $t=1$.
}}
\label{fig1}
\end{figure}

Using the Eqs. (\ref{ctrww}) and (\ref{ctrwalb}) we have the following asymptotic expansion in Laplace-Fourier space
 \begin{eqnarray}
 \psi(k,s)=1-( \tau s)^{\alpha}-k^2 \frac{\langle x^2 \rangle}{2},
 \label{assintotic1}
 \end{eqnarray}
in which $0<\alpha<1$. Replacing the Eq. (\ref{assintotic1}) in Eq. (\ref{distributionctrw}) obtain 
\begin{eqnarray}
p(k,s)=  \frac{s^{\alpha-1}p(k,0)}{s^{\alpha}+\mathcal{K}_{\alpha} k^2}, \qquad \mathcal{K}_{\alpha}= \frac{\langle x^2 \rangle}{2\tau^{\alpha}},
\label{walk1}
\end{eqnarray}
by performing the inverse Laplace-Fourier transforms we obtain the following equation
\begin{eqnarray}
\frac{1}{\Gamma(1-\alpha)}\frac{\partial}{\partial t}\int_0^t dt' \frac{p(x,t')}{(t-t')^{\alpha}}=\mathcal{K}_{\alpha}\frac{\partial^2}{\partial x^2}p(x,t),
\end{eqnarray}
which can be written as follows 
\begin{eqnarray}
_0^{RL}\textbf{D}_t^{\alpha} p(x,t')=\mathcal{K}_{\alpha}\frac{\partial^2}{\partial x^2}p(x,t),
\label{difusao1}
\end{eqnarray}
in which  
\begin{eqnarray}
_0^{RL}\textbf{D}_t^{\alpha}f(t)=\frac{1}{\Gamma(1-\alpha)}\frac{\partial}{\partial t}\int_0^t dt' \frac{f(t')}{(t-t')^{\alpha}},
\end{eqnarray}
is the fractional derivative of Riemann-Liouville \cite{podlubny1998fractional}. For more details on the fractional calculus see the reference \cite{podlubny1998fractional}. 
The solution (\ref{difusao1}) is given by
\begin{eqnarray}
p(x,t)&=&\frac{1}{\sqrt{4  \mathcal{K}_{\alpha}t^{\alpha}}}
{\mbox {\large{H}}}_{1,1}^{1,0} \left[ \frac{|x|}{\sqrt{\mathcal{K}_{\alpha}t^{\alpha}}} ~
\Bigg|_{ \left(0,\;  1 \right) }^{\left(1-\frac{\alpha}{2},  \alpha \right)}   \right],
\label{fracdifusion}
\end{eqnarray}
in which H is the Fox function \cite{fox}. If $\alpha =1$, the solution becomes Gaussian and the fractional diffusion equation becomes a usual. In fact, the result shown by Eq. (\ref{fracdifusion}) is related to a generalised distribution directly linked to the concept of non-integer derivative. The figure \ref{fig1} exemplifies the behaviours of Eq. (\ref{fracdifusion}). In particular, the fractional derivative in time is due to the link of the distribution of waiting times associated with power-law type distributions. For example, if we assume a generalised exponential function to waiting time distribution we obtain more complex models  \cite{sandev2018continuous}. Actually,  there is a big class of waiting-time distributions that are very important in physics \cite{maike2,emilia2018subordination}.

We derive a result that is a direct consequence of the fractional derivative in Eq. (\ref{difusao1}) considering that the particle has an initial state located at the origin, that is, $p(x,0)=\delta(x)$, thereby $p(k,0)=1$. As a consequence of the distribution to be symmetrical, we have $\langle x \rangle =0$. Therefore, we may calculate $\langle x^2 \rangle $ to associate this formalism with the anomalous diffusion. It is known that
\begin{eqnarray}
\langle x^2 \rangle (s) =  -\frac{\partial^2 \ }{\partial k^2} \mathcal{F}\lbrace p(x,s)  \rbrace \Bigg|_{k=0},
\end{eqnarray}
by using the Eq. (\ref{walk1}), we have
\begin{eqnarray}
\langle x^2 \rangle (s) = \left[ \frac{2 \mathcal{K}_{\alpha} s^{\alpha -1}}{\left(k^2 \mathcal{K}_{\alpha}+s^{\alpha }\right)^2}-\frac{8 k^2 \mathcal{K}_{\alpha}^2 s^{\alpha -1}}{\left(k^2 \mathcal{K}_{\alpha}+s^{\alpha }\right)^3} \right]\Bigg|_{k=0},
\end{eqnarray}
to $k=0$ and performing the inverse Laplace transform, we obtain
\begin{eqnarray}
\langle x^2 \rangle=\frac{2  \mathcal{K}_{\alpha} t^{\alpha }}{\Gamma (\alpha +1)},
\end{eqnarray}
since $\alpha$ is restricted to the range $0<\alpha <1$. This system type corresponds to a subdiffusive dynamics.

\subsection{The walker associated with L\'evy flights}

\begin{figure}[h]
\centering
\includegraphics[scale=1.0]{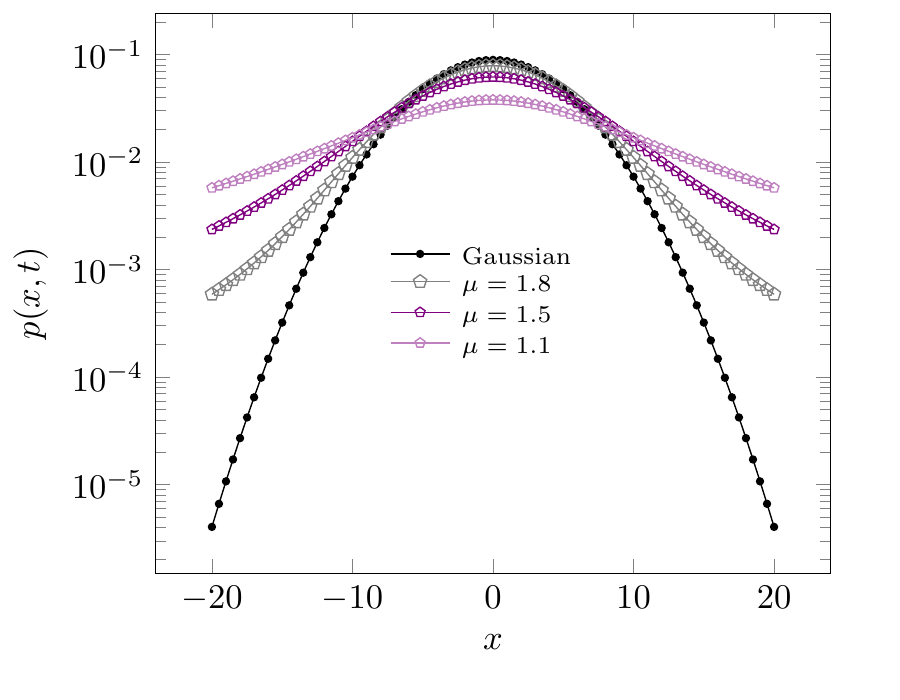}
\caption{\small{The figure present the  probability distribution for different values of $\mu$ parameter.  We consider $\mathcal{K}_{\mu}=10$ and $t=1$.
}}
\label{fig2}
\end{figure}

In this subsection, we will emphasise a kind of equation that takes into account the  divergent second moment to distribution, and these distributions are known as distributions of L\'evy, which are related to the so-called L\'evy flight \cite{levy,zaburdaev2015levy}. In this direction, the first moment of distribution $w(t)$ is considered finite, and the second moment in the infinite step length distribution. Considering the following form
\begin{eqnarray}
\lambda(k)=e^{-\sigma^{\mu}|k|^{\mu}} \sim 1-\sigma^{\mu}|k|^{\mu},
\label{ww}
\end{eqnarray}
in which $1<\mu <2$. Using the $w(s)\sim 1- \tau s $ and Eq. (\ref{ww}) in (\ref{distributionctrw}) we  may obtain
\begin{eqnarray}
p(k,s)=  \frac{p(k,0)}{s+\mathcal{K}_{\mu} |k|^{\mu}}, \qquad \mathcal{K}_{\alpha}= \frac{\sigma^{\mu}}{\tau},
\label{walk221}
\end{eqnarray}
simplifying this equation
\begin{eqnarray}
sp(k,s)-p(k,0)=-\mathcal{K}_{\mu}|k|^{\mu}p(k,s),
\end{eqnarray}
performing the inverse Laplace transform, we obtain
\begin{eqnarray}
\frac{\partial}{\partial t}p(k,t)=-\mathcal{K}_{\mu}|k|^{\mu}p(k,t),
\label{difusão12}
\end{eqnarray}
 we can define another fractional derivative. The Riesz-Feller proposal \cite{mainardi2001fundamental}, which can be written
\begin{eqnarray}
\frac{\partial^{\mu} f(x)}{\partial |x|^{\mu}}=\frac{\Gamma(1+\mu)}{\pi}\sin\left(\frac{\mu \pi}{2}\right)\int_0^{+\infty}d\xi \frac{f(x+\xi)-2 f(x)-f(x-\xi)}{\xi^{1+\mu}},
\end{eqnarray}
the Fourier transform of this derivative is expressed as follows
\begin{eqnarray}
\mathcal{F} \left\lbrace  \frac{\partial^{\mu} f(x)}{\partial |x|^{\mu}} \right\rbrace = -|k|^{\mu}f(k),
\label{13}
\end{eqnarray}
thus, the inversion of Eq. (\ref{difusão12}), the result is given by
\begin{eqnarray}
\frac{\partial}{\partial t}p(x,t)=\mathcal{K}_{\mu}\frac{\partial^{\mu} }{\partial |x|^{\mu}} p(x,t).
\end{eqnarray}
Now we want to find the solution of this fractional equation, with the initial condition is given by a delta function located at the origin, i.e.  $p(k,0)=1$. We may perform the inverse Laplace and Fourier transforms in Eq. (\ref{walk221}), we have
\begin{eqnarray}
p(x,t)=\frac{1}{2\pi}\int_{-\infty}^{+\infty}dk \varphi(k,t) e^{-i k x}, \qquad \varphi(k,t) =e^{-\mathcal{K}_{\mu}|k|^{\mu}t},
\label{levywalker}
\end{eqnarray}
the structure of $p(x,t)$ combined with the $\varphi$, characterises a L\'evy  distribution. The exact solution to integration the Eq. (\ref{levywalker}) is the following
\begin{eqnarray}
p(x,t)=\frac{1}{\mu |x|}
{\mbox {\large{H}}}_{2,2}^{1,1} \left[ \frac{|x|}{(\mathcal{K}_{\mu}t)^{\frac{1}{\mu}}} ~
\Bigg|_{\left(1,\;  1 \right), \left(1,\;  \frac{1}{2} \right) }^{\left(1, \frac{1}{\mu} \right),\left(1,  \frac{1}{2} \right)}  \right],
\end{eqnarray}
we can write the H-fox function in terms of L\'evy distribution, as follow
\begin{eqnarray}
p(x,t)=\frac{1}{(\mathcal{K}_{\mu}t)^{\frac{1}{\mu}}}
L_{\mu}\left[ \frac{|x|}{(\mathcal{K}_{\mu}t)^{\frac{1}{\mu}}} \right],
\label{levyevolution}
\end{eqnarray}
for $\mu=2$ we retrieve the Gaussian distribution, to $\mu=1$  we have another particular case that corresponds to the distribution of Cauchy
\begin{eqnarray}
p(x,t)= \frac{1}{\pi \mathcal{K}_1 t}\frac{1}{1+\frac{x^2}{\mathcal{K}_1^2 t^2}}.
\end{eqnarray}
The figure \ref{fig2} exemplifies the behaviours of Eq. (\ref{levyevolution}).

Various combinations can be made, and this type of formalism has been used to model various systems in physics. Examples of these are: diffusion of charges in neurons  \cite{neuroneo}, scattering patterns in the light spectrum \cite{luz1}, observation of anomalous diffusion and fractional self-similarity in one dimension \cite{sagi2012observation}, theory of fractional L{\'e}vy kinetics for cold atoms diffusing in optical lattices \cite{kessler2012theory}, aging renewal theory and application to random walks \cite{schulz2014aging}. In the references  \cite{Maike2017anomalous,maike1,maike2} we approaches  issues related to fractional diffusion.

In the last decades, the applicability of fractional equations (derived and integral) has been extensively investigated in several contexts that present anomalous diffusion. Ralf Metzler and Joseph Klafter reported a series of results in \cite{fractionalkramers1}, between these results, are present approaches on the Fokker-Planck equations and fractional Kramers, demonstrating analytically that the second moment of these equations present an anomalous diffusive behaviour. In \cite{fractionalkramers2},  Barkai and Silbey presented a version of the fractional Klein-Kramers equation in which they demonstrate an initial behaviour ($t \rightarrow 0$) ballistic, i.e., $\langle x^2(t) \rangle \sim t^2$ and for a long time the behaviour is super diffusive $\langle x^2(t) \rangle \sim t^{2-\alpha}$ in which $0<\alpha<1$. In this direction, the article \cite{PhysRevLett.96.230601} presents an extension of the concept of continuous time random
walks to position-velocity space. The authors derived a new fractional equation of the Kramers-Fokker-Planck type. The result is a series of anomalous in diffusive behaviour.

\section{Non-linear equations and the generalised walker}
\label{sec3}

Non-linear equations in the context of diffusion were extensively investigated as generalisations of the Fokker-Planck equation \cite{nonlinear,lenzi2,lisa1}. Just as the Einstein equation for diffusion has a mathematical structure identical to the heat equation, these nonlinear diffusion equations have similarities to equations that were already known. In this section, we mention such non-linear forms.

Considering a conservative system in which mass transport occurs, the density $\rho$ depends on the variables in time and space, it is possible to demonstrate that the dynamics are governed by the differential equation
\begin{eqnarray}
\frac{\partial \rho \phi}{\partial t}  = - \frac{\partial q\rho}{\partial x},
\end{eqnarray}
in which $q$ is related to the flow and $\phi$ the porosity of the medium. Using Darcy's Law \cite{darcy}, 
\begin{eqnarray}
q= -\frac{\kappa}{\mu}\frac{\partial \mathcal{P}}{\partial x},
\label{porous}
\end{eqnarray}
in which $\partial_x \mathcal{P}$ is the pressure gradient, $\kappa$ is the permeability and $\mu$ the viscosity of the fluid. Considering that the pressure has a non-linear shape with the density in porous media, there are several ways to approach this type of non-linearity. For example, we will use the power $\mathcal{P} \sim \rho^{\gamma-1}$, so the Eq. (\ref{porous}) can be written as
\begin{eqnarray}
\frac{\partial \rho}{\partial t}  = (constant) \times \frac{\partial^2 \rho^{\gamma}}{\partial x^2}. 
\end{eqnarray}
This equation is known as the transport equation in porous media \cite{vazquez2007porous,vazquez2006smoothing}, several extensions were investigated in fluid dynamics. Today, there are  numerical  \cite{netz1995computer,voigtmann2009double} and experimental  \cite{teixeira2007dynamic,fatin2004size} evidences that have related the diffusion in porous media with anomalous diffusion.

In 1995, Tsallis and Buckman investigated a nonlinear form for the Fokker-Planck equation which has the same mathematical structure as the porous media equation. The Tsallis-Buckman's work was entitled \textit{"Anomalous diffusion in the presence of external forces: exact time-dependent solutions and entropy"} \cite{tsallis1}. The proposal was
\begin{eqnarray}
\frac{\partial }{\partial t} [p(x,t)]^{\mu}=-\frac{\partial }{\partial x}[F(x) p(x,t)^{\mu}] + D \frac{\partial^2 }{\partial x^2} [p(x,t)]^{\nu},
\label{nontsallisequation}
\end{eqnarray}
this equation have a similar form to the equation for porous media \cite{spohn,logarithmic,schwammle2008q,mendes2008unified} for the case $F(x)=0$, suggesting that the mass flow caused by the pressure difference is analogous to the probability flow with non-linear terms. They presented the solution to the case where the force is given by $F(x)=k_1-k_2 x$ and demonstrate that the solution can be found by using the \textit{ansatz} given by
\begin{eqnarray}
p(x,t)= \frac{[1-\beta(t)(1-q)[x-x_M(t)]^2]^{\frac{1}{1-q}}}
{\mathcal{Z}_q}.
\label{ansatzq}
\end{eqnarray}
As a solution they find the general relation between the parameters, given by $q=1+\mu-\nu$, and the following solutions for $x_M(t)$, $\beta(t)$ and $\mathcal{Z}$,
\begin{eqnarray}
\frac{\beta(t)}{\beta(0)}=\Big[ \frac{\mathcal{Z}_q(0)}{\mathcal{Z}_q(t)} \Big]^{2\mu}, \qquad x_M(t)=\frac{k_1}{k_2}+\Big[ x_M(0) -\frac{k_1}{k_2}\Big]e^{-k_2 t},
\end{eqnarray}
and
\begin{eqnarray}
\mathcal{Z}_q(t)=\mathcal{Z}_q(0)\Big[ \Big(1-\frac{1}{K_2}\Big) e^{-\frac{t}{\tau}} + \frac{1}{K_2}\Big]^{\frac{1}{\mu+\nu}},
\end{eqnarray}
in which
\begin{eqnarray}
K_2=\frac{k_2}{2\nu \beta(0) D \mathcal{Z}_q(0)^{\mu-\nu}}, \qquad \tau = \frac{\mu}{k_2 (\mu+\nu)},
\end{eqnarray}
assuming the case where $k_2=0$ and the relationship $\beta = 1/2\sigma^2$, such that $\sigma$ is the variance of the distribution, as a consequence, we obtain the following expression
\begin{eqnarray}
\frac{1}{\beta(t)} \propto [\mathcal{Z}_q(t)]^{2\mu} \propto t^{\frac{2\mu}{\mu+\nu}}.
\end{eqnarray}
This result recovers the Brownian motion for the usual Fokker-Planck equation, that is, $\mu=\nu=1$,  that implies $\frac{1}{\beta(t)}\propto t$. If $\nu/\mu >1$ the system is subdiffusive and $\nu/\mu<1$ is subdiffusive, taking into account that $0<\nu/\mu <+\infty$.

The generalisation of the Gaussian function in Eq. (\ref{ansatzq}) is given by 
\begin{eqnarray}
e_q(-x^2)=[1-(1-q)x^2]^{\frac{1}{1-q}},
\end{eqnarray}
which is commonly known as $q$--Gaussian.
Just as the Brownian motion can be associated with the central limit theorem. The $q$--Gaussian distribution was associated with a generalisation of the central limit theorem \cite{tsallis2}. On the other hand the  $q$-exponential function has had several implications for diffusive phenomena as well as for other contexts in physics. In particular, Tsallis and others have used this formalism to find a class of solutions for nonlinear versions of the Schr\"{o}dinger, Klein-Gordon, and Dirac  equations \cite{tsallis3,tsallis4}.

The  q-exponential \textit{ansatz} was presented in literature in 1988 by physicist C. Tsallis in Ref. \cite{tsallis}. In the article, he proposed a non-additive entropic form 
\begin{eqnarray}
S_q=k\frac{1- \displaystyle\sum_{i=1}^W p_i^q}{q-1}, \qquad q\in \mathbf{R},
\label{tsallisentropy}
\end{eqnarray}
in which $\sum_{i=1}^W p_i=1$. The Eq. (\ref{tsallisentropy}) has many applications in physics and complex systems \cite{tsallis2017foundations,tsallis2016approach,tsallis2009computational,picoli2003q,picoli2009q,mendes2001renormalization,Nonextensive1,tsallis2017economics}.
The theory he formulated is known as non-extensive or generalised statistical mechanics, since to  $q\rightarrow 1$ the Tsallis entropy recovers the Boltzmann-Gibbs entropy, i.e. $\lim_{q \rightarrow 1}S_q = S_{BG}$. The proposed entropy brings with it a generalisation of Boltzmann factor and of the dynamics of Brownian motion. Here, it is necessary to remember that the  Tsallis-statistic can be connected with Superstatistic theory \cite{beck2003superstatistics}

The \textit{ansatz} (Eq. (\ref{ansatzq})) proposed by Tsallis is a consequence of the generalisation of Boltzmann factor \cite{temperatura}, i.e. $e_q^{-\beta E}$, in that in the limit $q\rightarrow 1$ the Eq. (\ref{ansatzq}) assumes a Gaussian form. In fact, the walker may be called \textit{generalised walker} because it is linked to non-extensive statistical mechanics. The connection between the Tsallis statistic and nonlinear Fokker-Planck equation (FPE) is deeper than the  Boltzmann factor. It was proved through the H--theorem 
 in works \cite{htheorem,dos2017nonlinear} to nonlinear FPE, and in Refs. \cite{htheorem1,chavanis2008nonlinear} to nonlinear Klein-Kramers equation. It means that, in works \cite{chavanis2008nonlinear,htheorem,PhysRevE.86.061136}, the authors use the H-theorem to  present how different class of nonlinear Fokker-Planck equation implies  generalised entropies (Tsallis, R\'enyi, Kaniadakis, etc) or vice versa.

As presented in the section \ref{sec1} of this chapter, we discuss an important work associated with the anomalous diffusion processes of particles suspended in turbulent currents. In \cite{lenzi3}, Malacarne et al., investigated a series of solutions for the nonlinear diffusion equation with the power law type diffusion coefficient, i.e., $K(r)=Dr^{-\theta}$. These equations can be summarised as follows
\begin{eqnarray}
\frac{\partial p}{\partial t}=\nabla K(r) \nabla p^{\nu},
\end{eqnarray}
in which the Laplacian is written as follows
\begin{eqnarray}
\nabla K(r) \nabla = r^{-(d-1)}\frac{\partial}{\partial r} Dr^{d-1-\theta} \frac{\partial}{\partial r},
\end{eqnarray}
 for $\theta=0$ the Laplacian assumes the form $ d $-dimensional. The solution found by the authors is a generalisation of nonlinear walker to a heterogeneous system. They obtained an analytic expression for the second moment, given by the following form
 \begin{eqnarray}
 \langle r^2 \rangle \propto t^{\sigma}, \qquad \textnormal{in which} \qquad \sigma=\frac{2}{2+\theta+ d(\nu-1)},
 \end{eqnarray}
 for $\nu=1$ we have $ \langle r^2 \rangle \propto t^{\frac{2}{2+\theta}}$. In addition, considering   $\theta=-\frac{4}{3}$ we recover the case proposed by Richardson in Eq. (\ref{richarsonmsd}). The consideration $\nu=1$ shows that the nonlinear case has a class of solutions in which MSD is invariant from the point of view of dimensions (parameter $ d $). In the case that $\theta=(1-\nu)d$, the system has a usual Brownian behaviour, if $ \theta > (1-\nu)d$ the system is subdiffusive and if $ \theta <(1- \nu) d $ the system is super-diffusive.
 
\begin{figure}[h]
\centering
\includegraphics[scale=1.0]{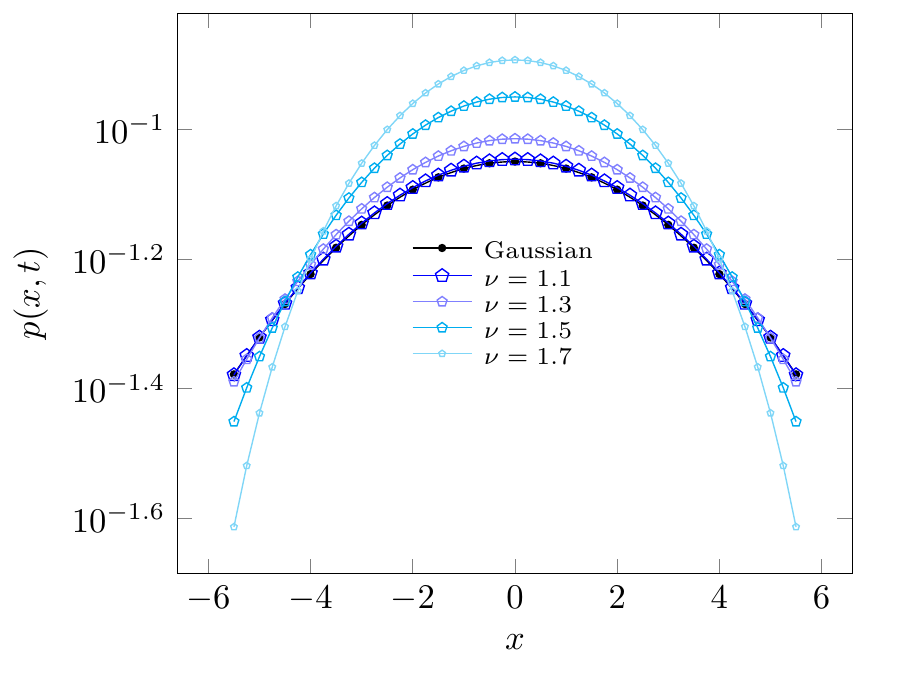}
\caption{\small{The figure present the  probability distribution for different values of $\nu>1$ parameter.  We consider $D=10$ and $t=1$.
}}
\label{fig3}
\end{figure}

\begin{figure}[h!]
\centering
\includegraphics[scale=1.0]{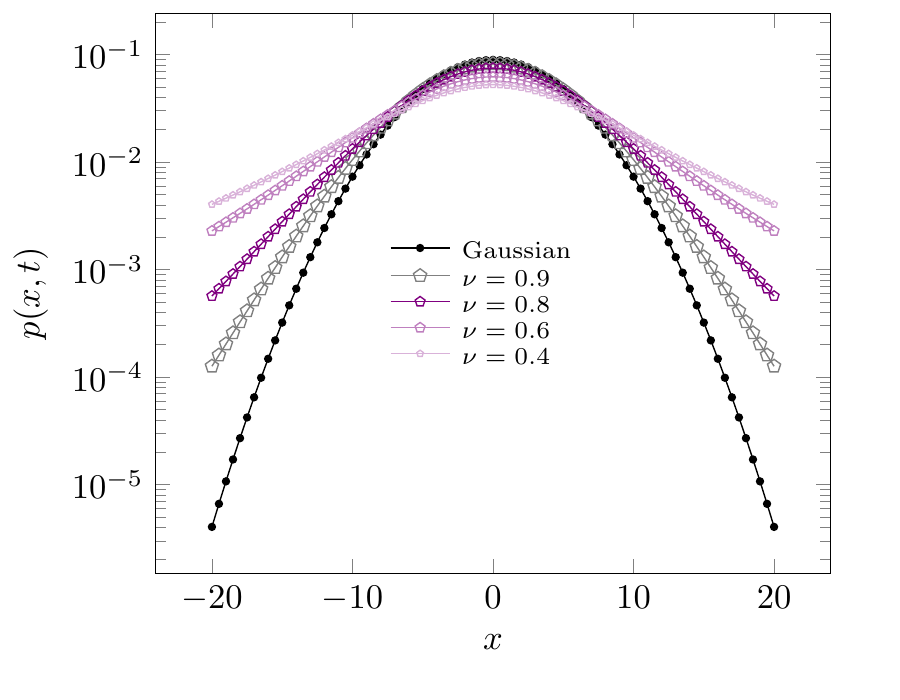}
\caption{\small{The figure present the  probability distribution for different values of $\nu<1$ parameter.  We consider $D=10$ and $t=1$.
}}
\label{fig4}
\end{figure}

 The figures \ref{fig3} and \ref{fig4} exemplify the behaviours of Eq. (\ref{nontsallisequation}) to $F(x)=0$, with normalisation factor presented in Ref. \cite{lenzi3}.

Therefore, there are several non-linear formulations in which the diffusion equation has been investigated \cite{lenzi4,plastinodifusao,Nonextensive1,lenzi2013solutions11gb,fuentes2008computing}. In particular, Tsallis's proposal was used to investigate quantum entangled  \cite{lenzi2013time11gb},  nonextensive scaling law in confined granular media \cite{combe2015experimental}, standard map \cite{tirnakli2016standard}, anomalous diffusion in a long-range Hamiltonian system \cite{moyano2006diffusive}, dynamics of normal and anomalous diffusion in nonlinear Fokker-Planck equations \cite{schwammle2009dynamics}, normal and tumoral melanocytes exhibit $q$-Gaussian random search patterns  \cite{da2014normal}.

\section{Stochastic equations and correlated walker}
\label{sec4}

\par The Langevin equation was generalised by Kubo and Mori for a class of correlated noises \cite{kubo,mori}, which made the equation known by the generalised Langevin equation (GLE). Their description incorporate the fractional formalism made by Mandelbrot, and resulted in a generalisation of the  fluctuation-dissipation theorem \cite{mandelbrot}. On the other hand, GLE has been used to describe systems out of equilibrium \cite{langevin3} and sub-diffusive processes of a single molecule of protein \cite{langevin2}.

Mandelbrot and Van Ness extended the Brownian motion by proposing a fractional Gaussian noise \cite{mandelbrot}, which is written as follows 
\begin{eqnarray}
B_H(t)&=&\frac{1}{\Gamma(H+\frac{1}{2})}\Bigg[\int_0^t (t-\tau)^{H-\frac{1}{2}} dB(\tau)+\int_{-\infty}^0 [(t-\tau)^{H-\frac{1}{2}}  \nonumber \\
&-&  (-\tau)^{H-\frac{1}{2}}] dB(\tau) \Bigg],
\label{varianciaH}
\end{eqnarray}
in which $H$ is the Hurst exponent, the Gaussian noise is recovered to the value $H=\frac{1}{2}$,  for other values between $0<H<1$ the noise incorporates anomalous effects, and for this reason is known as fractional noise. The variance of Eq. (\ref{varianciaH}) is given by $2D_Ht^{2H}$, in which $D_H=\Gamma(1-2H)\cos[H\pi]/(2H\pi)$. From these properties one can determine the correlation between noise, which for Mandelbrot noise can be written as
\begin{eqnarray}
\langle \xi(t)\xi(t') \rangle = 2D_H H(2H-1)|t-t'|^{2H-2}, \qquad t, t' >0
\label{xi2}
\end{eqnarray}
in which $\langle \xi(t) \rangle =0$.
\par The fractional Langevin equation is given by \cite{deng2009ergodic}
\begin{eqnarray}
m\frac{d^2}{dt^2}x(t)=- \varphi \int_{0}^t dt'  \frac{1}{|t-t'|^{2-2H}} \frac{d}{dt'}x(t') \ + \overline{\eta} \cdot \xi(t), 
\label{langecap2}
\end{eqnarray}
in which
\begin{eqnarray}
\overline{\eta}=\sqrt{\frac{k_BT \varphi }{2D_H H(2H-1)}},
\label{eta}
\end{eqnarray}
the Eq. (\ref{langecap2})  shows that the noise is correlated with the friction term through the fluctuation-dissipation theorem. In particular, the Eq. (\ref{langecap2}) can be related to the formalism of the fractional calculus as follows
\begin{eqnarray}
m\frac{d^2}{dt^2}x(t)=-\varphi \   _a^C\textbf{D}_t^{\alpha} x(t') \ + \overline{\eta} \cdot \xi(t), \qquad \alpha=2-2H.
\label{langev}
\end{eqnarray}
The fractional derivative $_a^C\textbf{D}_t^{\alpha}$ is of the Caputo type, defined as follows
\begin{eqnarray}
_a^C\textbf{D}_t^{\alpha}f(t)=\frac{1}{\Gamma(n-\alpha)}\int_{a}^{t}dt' \frac{1}{(t-t')^{\alpha+1-n}}\frac{d^n}{dt'^n}f(t'),
\end{eqnarray}
in which $n-1<\alpha <n$ \cite{podlubny1998fractional}.

A simplified version of the Eq. (\ref{langecap2}) occurs in the case when $m \rightarrow 0$ and in the absence of correlation in the friction term, which implies $ \dot{x}(t)= \varphi^{-1}\eta \cdot \xi(t)$ (the model A in Ref. \cite{deng2009ergodic}), and has as solution
\begin{eqnarray}
x(t)=\frac{\overline{\eta}}{\varphi}\int_0^{t}\xi(t')dt',
\end{eqnarray}
the ensemble average of $x^2$ can be calculated as follows
\begin{eqnarray}
\langle x^2 \rangle = \Big(\frac{\overline{\eta}}{\varphi}\Big)^2 \int_0^{t}\int_0^{t''}dt'dt'' \langle \xi(t')\xi(t'') \rangle,
\label{x22}
\end{eqnarray}
using Eq. (\ref{xi2}) and (\ref{eta}) in (\ref{x22}), we have
\begin{eqnarray}
\langle x^2 \rangle = \frac{2 D_H k_B T}{\varphi}t^{2H}, \end{eqnarray}
which is related to anomalous diffusive processes.

The generalised Langevin equation by Kubo and Mori is written as follows
\begin{eqnarray}
m\frac{d^2}{dt^2}x(t)=-\varphi \int_{0}^t dt'  \zeta(t-t') \frac{d}{dt'}x(t') \ + \eta \cdot \xi(t), 
\label{langevg1}
\end{eqnarray}
which implies the generalised version of the fluctuation-dissipation theorem theorem (FDT) given by \cite{kubo}
\begin{eqnarray}
\langle \xi(t)\xi(t') \rangle \propto  \zeta(t-t'), \qquad t, t' >0.
\end{eqnarray}
This kind of generalisation made it possible for a class of non-Gaussian noises to be investigated and referred to as coloured noises. In particular, the FDT recovers the usual Brownian motion for the case where $\zeta(t)=\delta(t)$ (delta function). This case, applied in Eq. (\ref{langevg1}), implies a dynamics that corresponds to the Klein-Kramers equation \cite{risken}.

A class of stochastic walkers on the influence of mutiplicative noise can be written as \cite{risken}
\begin{eqnarray}
\frac{dx}{dt}=h(x(t),t)+g(x(t),t)\xi(t),
\label{ito}
\end{eqnarray}
that consists in the Ito-Langevin equation.
This equation was used in \cite{lisa1} to describe the stochastic movement of generalised walkers, as seen in the previous section. The Eq. (\ref{ito}) has the following correspondence with the Fokker-Planck equation
\begin{eqnarray}
\frac{\partial }{\partial t}p(x,t)=-\frac{\partial }{\partial x}[h(x,t)p(x,t)]+\frac{1}{2}\frac{\partial^2}{\partial x^2}[g^2(x,t)p(x,t)].
\label{difcap2}
\end{eqnarray}
The correspondence between formalisms allows us to perform a more complete analysis of the physical system. To exemplify this fact, let us consider the generalised dynamics of Richardson. Assuming that $g^2(x,t)=k |x|^{\gamma}$, we can write the Eq. (\ref{difcap2}) this way
\begin{eqnarray}
\frac{\partial }{\partial t}p(x,t) &=& \frac{\gamma}{2}\frac{\partial}{\partial x}[k |x|^{\gamma-1} p(x,t)]+\frac{k}{2}\frac{\partial}{\partial x}\Big[|x|^{\gamma}\frac{\partial}{\partial x}[p(x,t)]\Big] \nonumber \\
&-&\frac{\partial }{\partial x}[h(x,t)p(x,t)].
\end{eqnarray}
To recover the Richardson diffusion equation, i.e. Eq. (\ref{rich1}), considering  
\begin{eqnarray}
h(x,t)=k\frac{\gamma x^{\gamma-1}}{2},
\end{eqnarray}
we obtain an equation in Richardson sense, as follows
\begin{eqnarray}
\frac{\partial }{\partial t}p(x,t)=\frac{k}{2}\frac{\partial}{\partial x}\Big[x^{\gamma}\frac{\partial}{\partial x}[p(x,t)]\Big].
\label{Rich}
\end{eqnarray}
Therefore, we can write Eq. (\ref{Rich}) for the movement of particles in turbulent medium  as follows 
\begin{eqnarray}
\frac{dx}{dt}=k\frac{\gamma |x(t)|^{\gamma-1}}{2}+\sqrt{k |x(t)|^{\gamma}}\xi(t),
\label{ito2}
\end{eqnarray}
in this case $\xi$ is a Gaussian noise. We have established a connection between the diffusion equation and the stochastic Langevin equation for the diffusion proposed by Richardson for the turbulent diffusion \cite{richardson1}.

The Eq. (\ref{ito}) can represents a class of coloured noise equations. Suppose that the noise $\xi(t)$ is not white and is generated by another equation, which is represented here by
\begin{eqnarray}
\frac{d\xi}{dt}=-\frac{1}{\tau} \xi(t)+ \frac{c}{\tau} \overline{\xi}(t),
\label{estocasticacap2}
\end{eqnarray}
in which $\overline{\xi}(t)$ is given by white noise, in which 
\begin{eqnarray}
\langle \overline{\xi}(t')\overline{\xi}(t) \rangle =\delta(t-t').
\end{eqnarray}
Whereas $\langle \xi(t) \rangle =0$, we have
\begin{eqnarray}
\langle \xi(t')\xi(t) \rangle = \frac{c^2}{2\tau}  e^{-\frac{| t-t'|}{\tau}},
\end{eqnarray}
if $\tau \rightarrow 0$ the Eq. (\ref{estocasticacap2}) assumes the form $\xi= c \overline{\xi}$, so the Eq. (\ref{ito}) is written in terms of white noise.

This formalism closes the last analytic form we will address in this review article. All methods in this section have simplicity in describing  of the mean quadratic displacement associated with a stochastic process, so they gain more and more space in current research.

\section{Brief discussion and some considerations}
\label{sec5}

The objective since the text was to present the different ways of approaching anomalous diffusive processes, which covers a diversity of problems. The mechanisms that can lead to anomalous diffusion are the most diverse. Mentioning some examples, we have: systems with traps, diffusion in fractal structures, non-locality, interactions, temporal memory. In order to the reader may acquire a general understanding of  the analytical approaches to anomalous diffusion, see the table \ref{table1}.

Among a number of applications that the methods in this review can be applied, we wish to mention some examples and new directions of these methods in the description of anomalous diffusion phenomena.
\begin{enumerate}
    \item   The first formalism presented in this review was related to the fractional walker, this method has a great focus of investigation in the present day, because daily new techniques, theorems and theories continue to be developed by the mathematicians who investigate the fractional calculus \cite{de2019fractional,Baleanu1,Baleanu2,Baleanu3,Baleanu4,Baleanu5,duan1,duan2,atangana1,atangana2,atangana3,atangana4,atangana5}. With all this progress of the fractional calculus, physics advances together, since it allows the modelling of a series of interesting problems in physics, such as diffusion equation with tempered derivatives \cite{tempered1,tempered2,tempered3,tempered4,tempered5,tempered6}, memory systems  \cite{hristov1,hristov2,hristov3,hristov4,hristov2017derivation}, non-homogeneous systems \cite{maike2,heterogeneous1,heterogeneous2,heterogeneous3}, etc \cite{hernandez2017self,singh2018analysis,das2018time}. One of the most important roles of fractional calculus in physics has been to introduce ever more sophisticated memory kernels, which capture more precisely the processes observed in the real world \cite{sun2018new}. \\
    
    \item The second formalism presented was related to a non-linear diffusion equation. We show that these equations gain a foundation when they are involved with the proposal of non-extensive statistical mechanics \cite{tsallis}. Nowadays, the Tsallis statistic \cite{tsallis} and nonlinear diffusion have assumed important roles in the application of more subtle problems in thermodynamics, such as black holes \cite{tsallis2013black,caruso2008nonadditive}, generalised forms of H-theorem  \cite{htheorem1,dos2017nonlinear,htheorem3}, financial market \cite{michael2003financial,queiros2007nonextensive}, and many other systems \cite{Nonextensive1,Nonextensive2,Nonextensive3}. Here, it is worth mentioning to the reader that from the point of view of the H theorem \cite{htheorem} the porous media equations may imply the Tsallis entropy. However, there are certain non-linear combinations of the diffusion equation that can imply other entropic forms such as R\'enyi, Kaniadakis, etc. Mathematically, the inverse problem is also valid. From generalised entropies, we can obtain  generalised dynamics. However, a dynamic usually implies an entropy. Therefore, the Tsallis entropy would be a consequence of non-linearity in the Fokker-Planck equation  \cite{htheorem,chavanis2008nonlinear,entropyproduction1,PhysRevE.86.061136}. \\
    
    \item The third formalism addressed by us was the generalised Langevin equation. At present this technique has been extensively investigated in ergodicity breaking \cite{deng2009ergodic,ergodic2,ergodic3},  superstatistic \cite{super1,super2}, among others \cite{noise1,cheng2018moderate,baldovin2018langevin,ferreira2012analytical}. The idea of a noise introduced by Langevin is something that transcends physics and has become important even in financial systems \cite{kluppelberg2004fractional},  anomalous diffusion of telomeres in the nucleus of mammalian cells \cite{bronstein2009transient}, active walker model for the formation of human and animal trail systems \cite{helbing1997active}. Recently, Oliveira \textit{et al.} published a paper \cite{oliveira2019anomalous} that has a deep approach to the generalised Langevin equation. In addition, the authors introduced a general discussion about other formalisms that are essential to describe anomalous diffusive systems.
\end{enumerate}

\begin{table}[h!]
\scriptsize
\begin{tabular}{|c|c|c|c|}
\hline
\begin{tabular}[c]{@{}c@{}} \\ Diffusion \\ \\ \end{tabular}       & \begin{tabular}[c]{@{}c@{}} \\ Fractional \\ equations \\  \\ \end{tabular}                                         & \begin{tabular}[c]{@{}c@{}} \\ Nonlinear \\ equations \\   \\ \end{tabular}                                                                                             & \begin{tabular}[c]{@{}c@{}} \\ Stochastic \\ equations \\  \\   \end{tabular}                                                                                                            \\ \hline
\begin{tabular}[c]{@{}c@{}} \\ Some \\ applications \\ \\ \end{tabular}    & \begin{tabular}[c]{@{}c@{}}    \\ Self-similarity \cite{sagi2012observation}; \\ Cold atoms  \cite{kessler2012theory}; \\ Weak ergodicity \\  \cite{ergodic3,burov2010aging,bouchaud1992weak}; \\ Amorphous solids \cite{scher1975anomalous}; \end{tabular} &  \begin{tabular}[c]{@{}c@{}} \\ Porous media \cite{porousmediaflowMuskat,nonlinear}; \\ Thin Liquid films \\ under gravity \cite{buckmaster1977viscous}; \\
Radiative heat  \\
transfer \cite{larsen1980asymptotic}; \\  Surface dynamics  \cite{spohn1993surface};

\end{tabular}                                        &                               \begin{tabular}[c]{@{}c@{}}    { \centering Transient anomalous} \\ { diffusion cells \cite{bronstein2009transient}}; \\ Biomolecular \\ Folding \cite{satija2019generalized}; \\ 
Protein motion \cite{langevin2};
 \end{tabular}                                                                                \\ \hline
\begin{tabular}[c]{@{}c@{}} \\ Some \\ connections \\ \\ \end{tabular}          &   \begin{tabular}[c]{@{}c@{}} \\ Central limit \\ theorem \cite{bouchaud1990anomalous};      \\ CTRW \cite{ctrw}; \\ L\'evy flight  \cite{zaburdaev2015levy};  \\
Fractional variable-order \\ 
derivative \cite{yang2017new}; \end{tabular}                             &   \begin{tabular}[c]{@{}c@{}} \\ H-theorem \cite{htheorem}; \\ Generalised entropies \cite{chavanis2008nonlinear,gell2004nonextensive} \\ Superstatistics \cite{beck2003superstatistics,beck2004superstatistics};\\ Random Walks \cite{lenzi4}; \\ Ionised Plasma \cite{ebne2019transport}; \\ Radiation diffusion \cite{hristov3}; 
\end{tabular} &   
\begin{tabular}[c]{@{}c@{}} \\ Superstatistical  \cite{super1,sposini2018random};\\ Ergodic properties \cite{deng2009ergodic}; \\ Generalised correlations \\ functions \cite{sandev2017generalized,ferreira2012analytical};\\ Nonlinear systems \cite{lisa1}; \\

\end{tabular}

\\ \hline
\begin{tabular}[c]{@{}c@{}} Some\\ new\\ analytical\\ approaches\\ \\ \end{tabular}        &             \begin{tabular}[c]{@{}c@{}} \\  Prabhakar-difusion \cite{garra2014hilfer,physics1010005};  \\ Hristov-difusion \cite{sene2019analytical,sene2019solutions};   \\ Distributed-order \\  diffusion \cite{PhysRevE.92.042117}; \\ \\ \end{tabular}                                           &               \begin{tabular}[c]{@{}c@{}} \\
2-d Turbulence \cite{egolf2018tsallis}  \\ 
 Overdamped  \\  motion \cite{PhysRevE.98.012129}; \\  Confined interacting \\ particles \cite{Nonextensive1,htheorem1}; \\ Intermittent  motion  \cite{lenzi2017intermittent};  \\
\end{tabular}                             & \begin{tabular}[c]{@{}c@{}} \\  Anomalous diffusion \\ in the Ising model \cite{PhysRevE.98.012124}; \\ Superstatistical \cite{super1}; \\ Tempered Langevin \\ equation \cite{tempered4}; \\ \end{tabular} \\ \hline
\end{tabular}
\caption{A framework of some ideas}
\label{table1}
\end{table}

In the last decades, the anomalous diffusive processes have ceased to be restricted to analytic techniques. For the anomalous diffusion ceased to be a line of investigation only of physic and became a phenomenon observed in many fields of science. Today,  with the advancement of the computational approach, we can observe the anomalous diffusion from microorganisms \cite{leptos2009dynamics} to diffusion of particles across membrane \cite{coker2016membrane}. Analysing the trajectory of one stochastic system we can describe if the movement is classified as anomalous or not. Recently Pan Tan \textit{et al.}  made a combination of techniques, including the analytic, experimental and simulation, to describe the anomalous diffusion of water molecules around two biomolecules \cite{Pantan}. The techniques converge to a unique result, showing to the reader that the greater the number of techniques the greater the chances of unravelling some complex behaviour of nature.

\section*{Acknowledgements} 
M. A. F. dos Santos acknowledges the support of the Brazilian agency CNPq.







\end{document}